\documentclass[cernpreprint, texlive=2023, USenglish, texmf, orcidlogo]{atlasdoc}
 
\usepackage{atlaspackage}
\usepackage{placeins}
\usepackage{atlasbiblatex}
 
\usepackage{atlasphysics}

\graphicspath{{logos/}{figures/}}
 
\usepackage{ANA-HION-2021-08-PAPER-defs}
\usepackage{comment}

\AtlasTitle{
Measurement of the Sensitivity of Two-Particle Correlations in $pp$ Collisions to the Presence of Hard Scatterings
}

\AtlasAbstract{
A key open question in the study of multi-particle production in high-energy
\pp\ collisions is the relationship between the ``ridge''  --
observed azimuthal correlations between particles in the
underlying event that extend over all rapidities -- and hard or
semi-hard scattering processes.
In particular, it is not known whether jets or their soft fragments
are correlated with particles in the underlying event.
To address this question, two-particle correlations are measured
in \pp\ collisions at $\sqrt{s}=13$~\TeV\ using data collected by the
ATLAS experiment at the LHC, with an integrated luminosity of
$15.8$~$\mathrm{pb}^{-1}$, in two different configurations.
In the first case, charged particles associated with jets are excluded from the
correlation analysis, while in the second case, correlations are measured between
particles within jets and charged particles from the underlying event.
Second-order flow coefficients, $v_2$, are presented as a function of event
multiplicity and transverse momentum.
These measurements show that excluding particles associated with jets
does not affect the measured correlations.
Moreover, particles associated with jets do not exhibit
any significant azimuthal correlations with the underlying event,
ruling out hard processes contributing to the ridge.
}
 
\author{The ATLAS Collaboration}
 
\AtlasRefCode{HION-2021-08}
 
\PreprintIdNumber{CERN-EP-2023-036}

\AtlasJournalRef{Phys. Rev. Lett 131 (2023) 162301}
\AtlasDOI{10.1103/PhysRevLett.131.162301}

\hypersetup{pdftitle={ATLAS document},pdfauthor={The ATLAS Collaboration}}
 
\begin{document}
 
\maketitle

In heavy-ion collisions, two-particle correlations~(\tpc) in relative azimuthal
angle with large pseudorapidity~\cite{AtlasCoordinate} separation show distinct long-range
correlations~\cite{ALICE:2011svq,CMS-HIN-11-006,HION-2011-01,HION-2012-03,HION-2012-04,HION-2012-06,HION-2012-10,HION-2015-14,HION-2016-06,PhysRevC.66.034904,PhysRevC.93.051902}.
These long-range correlations are a simple manifestation of the single-particle
anisotropies, \vn, that originate from the hydrodynamic expansion of the
quark-gluon plasma (QGP) produced in these collisions.
The \vn\ are defined by parameterizing the azimuthal distribution of
produced particles as:
\begin{equation}
\frac{dN}{d\phi}\propto\left(1+2\sum_{n=1}^{\infty}\vn\cos(n(\phi-\Psi_{n}))\right),
\label{eq:single}
\end{equation}
where $\phi$ is the azimuthal angle of the particle momentum and \vn\ and $\Psi_{n}$
are the magnitude and phase of the $n^{\mathrm{th}}$-order anisotropy,
see Refs.~\cite{HION-2011-01,HION-2016-06} and references therein.
 
Because of their hydrodynamic origin in nucleus-nucleus (\NucNuc) collisions, such
long-range correlations were not expected  in smaller colliding systems
such as proton--nucleus~($p$+A) or proton--proton~(\pp) collisions, where collective
phenomena were not commonly expected to develop.
However, measurements by CMS showed the presence of such
long-range correlations, known as the ``ridge,'' in
high-multiplicity \pp collisions~\cite{CMS-QCD-10-002}.
Further investigations by ATLAS~\cite{HION-2015-09,HION-2015-14,HION-2016-01}
have demonstrated that these long-range correlations in \pp collisions are produced
from single-particle anisotropies similar to those in heavy-ion collisions.
These long-range correlations have been interpreted as evidence
of collective effects similar to those seen in heavy-ion collisions.
However, some authors have proposed that the ridge primarily results from
correlated production of partons in the presence of dense gluonic
initial states (i.e. the GLASMA)~\cite{Dumitru:2010iy,Dusling:2012iga,Gotsman:2017ifr,Dusling:2017aot,Altinoluk:2020wpf},
implying that much of the correlation structure associated with the ridge
should be associated with hard- or semi-hard scattering processes.
Previous measurements~\cite{HION-2017-11} have shown that the ridge is
unmodified in \pp\ collisions producing a $Z$ boson, but no
direct measurement in \pp\ collisions of the correlation
between jets or their fragments and the underlying event has yet
been performed, while such a correlation has been observed in \pPb\
collisions~\cite{HION-2018-20,ALICE:2022cwa}.
 
This \papertype\ presents \tpc\ measurements in \pp\ collisions at a
center-of-mass energy ($\sqs$) of 13~\TeV, using the ATLAS detector at the LHC\@.
The measurements are performed with two different particle-pair selections.
The first case explores correlations between tracks that are not jet constituents,
while the second case measures correlations between tracks
that are constituents of jets and tracks that are well-separated
from jets.
Similar measurements in \pPb\ collisions have shown significant non-zero
$v_2$ for low~\cite{ALICE:2022cwa} and high~\cite{HION-2018-20}
transverse momentum (\pt) particles generated in hard processes.
Correlations are also measured in events that are explicitly selected
by requiring the presence or absence of low-\pt\ jets.
These measurements can address whether or not the presence of jets affects the ridge,
and if the particles from jets exhibit azimuthal correlations with particles
from the underlying event and therefore contribute to the ridge.

The measurements presented here are performed using the ATLAS~\cite{PERF-2007-01}
inner detector~(ID), minimum-bias trigger scintillators~(MBTS), calorimeters
and the trigger and data acquisition systems~\cite{TRIG-2016-01}.
The ID records charged-particle trajectories within the pseudorapidity range
$|\eta| < 2.5$ using a combination of silicon pixel detectors including
the ``insertable B-layer''~(IBL)~\cite{ATLAS-TDR-2010-19,Abbott:2018ikt},
silicon microstrip detectors~(SCT), and a straw-tube transition radiation tracker (TRT),
all immersed in a 2\,T axial magnetic field~\cite{IDET-2010-01,AtlasCoordinate}.
The ATLAS calorimeter system consists of a liquid argon (LAr) electromagnetic
calorimeter covering $|\eta|<3.2$, a steel--scintillator sampling hadronic
calorimeter covering $|\eta|<1.7$, a LAr hadronic calorimeter covering
$1.5<|\eta|<3.2$, and two LAr electromagnetic and hadronic forward calorimeters
(FCal) covering $3.2<|\eta|<4.9$.
The ATLAS trigger system~\cite{PERF-2011-02} consists of a Level-1 (L1)
trigger implemented using a combination of dedicated electronics and
programmable logic, and a software-based high-level trigger (HLT).
An extensive software suite~\cite{ATL-SOFT-PUB-2021-001} is used in data simulation,
in the reconstruction and analysis of real and simulated data, in detector operations,
and in the trigger and data acquisition systems of the experiment.

The data were collected during Run\,2 of the LHC (2015--2018), with an average
collision rate per bunch crossing ($\mu$) of less than 3, and an
integrated luminosity of 15.8~\ipb.
The data used here were recorded using multiple minimum-bias, high-multiplicity,
and jet triggers, which are described in Ref.~\cite{SupplementaryMaterial}.
Additional offline requirements are imposed on the events selected by the triggers.
The events are required to have a reconstructed vertex with $|z|<100$~mm.
To suppress events with more than one \pp\ collision in the same bunch crossing,
events are required to have only one reconstructed vertex.
Pileup events where the vertices from multiple collisions are
sufficiently close such that they are reconstructed as a
single vertex are not removed by the one vertex requirement.
However, such merged events typically have a broader distribution
for the longitudinal impact parameter of tracks relative
to the vertex ($|z_0\sin(\theta)|$).
Such events are reduced by requiring
that the standard deviation of $|z_0\sin(\theta)|$ for all tracks
in an event is less than 0.25~mm.

The reconstruction and performance of tracks and primary vertices in the ID are
described in Refs.~\cite{Cornelissen_2008,Salzburger:2018442,ATL-PHYS-PUB-2015-006}.
The specific track selection criteria can be found in Ref.~\cite{SupplementaryMaterial}.
The track reconstruction efficiencies $\epsilon(\pT,\eta)$
are obtained using Monte Carlo (MC) generated events that are
passed through a \GEANT~\cite{Agostinelli:2002hh} simulation~\cite{SOFT-2010-01}
of the ATLAS detector and reconstructed using the procedures applied to the data.
The efficiency varies between 69\% and 87\% as a function of $\eta$ and \pt.

Jets used in this analysis are reconstructed using the anti-$k_t$
algorithm~\cite{Cacciari:2008gp} with a radius parameter of 0.4.
The inputs to jet reconstruction are ``particle flow objects'' as detailed in
Ref.~\cite{PERF-2015-09}.
Jets are calibrated to the hadronic scale using scale factors obtained from
MC simulations specifically derived for low-$\mu$ data.
Additional \textit{in situ} corrections~\cite{JETM-2018-05} are applied, which
account for differences in the jet response between the MC samples and data.
One issue in this analysis is that the modulation in the soft particles
in the event (Eq.~\eqref{eq:single}) biases the jet \pt\ in a manner that
depends on its orientation relative to the $\Psi_n$.
This affects the measurements of the correlations between
jet-fragments and the underlying event (UE) particles (discussed in detail below).
To mitigate this effect, instead of selecting jets based on their \pt,
selections are made on the following groomed quantity:
\begin{eqnarray}
p_{\mathrm{T}}^{\mathrm{G}}=\left|\sum_{\mathrm{constituents}}{\boldsymbol{p}_{\mathrm{T}}^{>\ 4\ \GeV}}\right|,
\label{eq:ptp}
\end{eqnarray}
where the sum runs over all the jet constituents with $\pt>4$~\GeV,
which considerably reduces the number of UE particles
within the jet, and makes this bias negligible, as shown in Ref.~\cite{SupplementaryMaterial}.

In previous ATLAS measurements of \tpcs\ in \pPb~\cite{HION-2012-13,HION-2013-04}
and \pp~\cite{HION-2015-09,HION-2016-01,HION-2017-11} collisions,
events were quantified by \nchrec: the total number of reconstructed
tracks with $\pt>0.4$~\GeV, passing the track selections discussed above.
In this analysis, a slight modification is made to ensure that the
event activity is not biased by the presence of jets,
and only reflects the soft multiplicity in the event.
The number of constituent tracks in jets with $\ptp\ > 15~\GeV$ is subtracted
from the measured multiplicity, and the corrected quantity, \ntrkreccorr,
is used to represent the event activity.
While counting the constituent tracks of jets,
the $\pt>4$~\GeV\ requirement is not imposed on the tracks.
Additionally, this correction is offset by the average number of UE tracks within the jet cone.
This offset is estimated by measuring the average number of tracks,
as a function of $\eta$ and $\phi$, that are in a $R=0.4$ cone in events
with similar multiplicity and trigger conditions.

In 2PC measurements, the distribution of particle-pairs
in relative azimuthal angle $\dphi=\phia-\phib$ are measured.
The labels $a$ and $b$ denote the two particles in the pair.
In evaluating the correlation functions, the tracks are weighted by the inverse of their reconstruction efficiency, $1/\epsilon(\pt,\eta)$.
To suppress short-range correlations, the particles are required
to have a pseudorapidity separation of $|\Delta\eta| > 2$.
In \pp\ collisions, back-to-back dijets also make a significant
contribution to the \tpcs.
To remove this contribution, a template-fit method~\cite{HION-2015-09,HION-2016-01,HION-2017-11}
is employed in which the measured \tpc\ is described by a fit having two components.
The first component accounts for the dijet contribution, $\cperi(\dphi)$,
which is measured using low-multiplicity events (called the ``peripheral reference'').
This analysis uses the \ntrkreccorr\ interval of 10--30 to build $\cperi$.
The second component accounts for the bulk contribution with a relative harmonic modulation, $\cridge(\dphi)$.
The 2PC can then be described as:
\begin{eqnarray}
C(\dphi) &=     & F\cperi(\dphi)+G\left(1+2\sum_{n=2}v_{n, n}\cos(n\dphi)\right) \nonumber\\
&\equiv& F\cperi(\dphi)+\cridge(\dphi), \label{eq:template}
\end{eqnarray}
where $F$ and \vnn\ are fit parameters and $G$ is fixed by the requirement that
the integrals of the fit and $C(\dphi)$ are equal.
The Fourier moments, \vnn, obtained from the template-fit quantify
the strength of the long-range correlation.
It is demonstrated in Refs.~\cite{HION-2015-09,HION-2016-01} that
the \vnn\ in \pp\ collisions obtained from Eq.~\eqref{eq:template}
factorize as $v_{n, n}(\pta, \ptb)=v_{n}(\pta)v_{n}(\ptb)$,
where \vn\ is the single particle anisotropy (Eq.~\eqref{eq:single}).
Thus, $\vn(\ptb)$ is obtained as $\vn(\ptb) = \vnn(\pta, \ptb) / \sqrt{\vnn(\pta, \pta)}$.

The tracks used in this analysis are categorized as follows:
those that are separated from all $\ptp>15~\GeV$ jets by at least one unit in $\eta$~\cite{HION-2018-20}
and having $0.5 < \pt\ < 4~\GeV$ are considered to be UE tracks (\UET);
tracks that are included as particle-flow constituents of jets having
$\ptp>40~\GeV$ (called \textit{trigger jets} henceforth)
are considered to be jet constituents (\JT).
Five classes of correlations are studied in this \papertype:
\begin{itemize}
\item \hh:      Standard 2PC~\cite{HION-2015-09, HION-2016-01} without applying any rejection of tracks around jets.
\item \All:     2PC where both tracks are \UET. About 14\% of \hh\ 2PC pairs are removed by the abovementioned rejection.
\item \NoJet:   2PC using events with no jets with $\ptp>15~\GeV$.
\item \WithJet: 2PC using events with at least one jet with $\ptp>15~\GeV$.
\item \hjet:    2PC performed between \UET\ and \JT.
\end{itemize}
These classes are not mutually exclusive.
Specifically, the \NoJet\ and \WithJet\ classes add up to the \All\ class.
The \hjet\ class has no overlapping particle-pairs with the ones in the
\All, \NoJet, and \WithJet\ classes.
The \hh\ class is identical to the measurements performed in the previous
ATLAS publications~\cite{HION-2015-09,HION-2016-01},
and is used as a reference with which other classes are compared.
 
For the \hjet\ case, additional requirements are imposed on the trigger jets
to avoid distortions of the 2PC.
They must have no other jet with $\ptp>15~\GeV$ within $\Delta R=\sqrt{\Delta\eta^2+\Delta\phi^2} = 1$
and they must have a balancing jet with $\ptp>15~\GeV$ and with $|\dphi| > 5\pi/6$.
The first requirement removes distortions of the 2PC at smaller \dphi\
while the second requirement ensures that fragments of the balancing jet are
excluded from \UET.

It may happen that some constituents of jets originate in the UE,
leading to a contribution of combinatorial pairs in the \tpc.
These combinatorial pairs, by construction, have the same correlation
as those where both the tracks are from the UE.
The contribution of such pairs is removed by the following technique.
For each event that contributes to the \hjet\ correlation, a separate
\tpc\ is made using another event with similar vertex position and multiplicity.
In this event, one track is picked from an $\eta$-$\phi$ region that is
within $R=0.4$ cone of the jet-axis and the other track is picked from
the same $\eta$ range as in the \hjet\ event.
This combinatorial \tpc\ is then subtracted from the \hjet\ \tpc.

Statistical uncertainties in the measured \tpcs\ are evaluated using a
bootstrapping procedure previously used in Ref.~\cite{HION-2018-27}.
Systematic uncertainties in the $v_2$ measurements are estimated by varying
different aspects of the analysis.
For the template-fit procedure, the \ntrkreccorr\ multiplicity range for
the peripheral reference selection was varied from the nominal 10--30 to
10--40 and 20--40~\cite{SupplementaryMaterial} and the change in the
$v_2$ values is included as a systematic uncertainty.
For the multiplicity dependence, this uncertainty for the $v_2$ is 0.01 (absolute)
for the \hjet\ class
and is typically within 2\% for the other classes.
This uncertainty is fully correlated across all multiplicity intervals and
is the dominant uncertainty for the \hjet\ class.
Uncertainties in the tracking efficiency are propagated into the measured $v_2$.
This uncertainty on the $v_2$ is less than 0.5\%, and is estimated by varying
the efficiency up and down within its uncertainties ($\sim\pm$3\%)~\cite{HION-2019-03},
and re-evaluating the $v_2$.
The systematic uncertainty due to non-primary tracks is estimated by varying
the selection criteria for transverse and longitudinal impact parameters,
resulting in a 0.5\% change in $v_2$.
The \tpc\ analyses often use event-mixing~\cite{HION-2011-01,HION-2016-06} to
estimate and correct the \tpcs\ for the detector's pair acceptance.
This correction is quite small, and the full effect of the correction is included
as a systematic uncertainty.
As discussed previously, the events used in this analysis are required
to have the standard deviation of $|z_0\sin(\theta)|$ for the tracks
in an event to be smaller than 0.25~mm, to reduce pileup.
Conservatively, the entire effect of this selection,
which varies with multiplicity but is typically within 1\%,
is taken to be a systematic uncertainty associated with pileup effects.

\begin{figure}[h]
\centering
\includegraphics[width=0.95\linewidth]{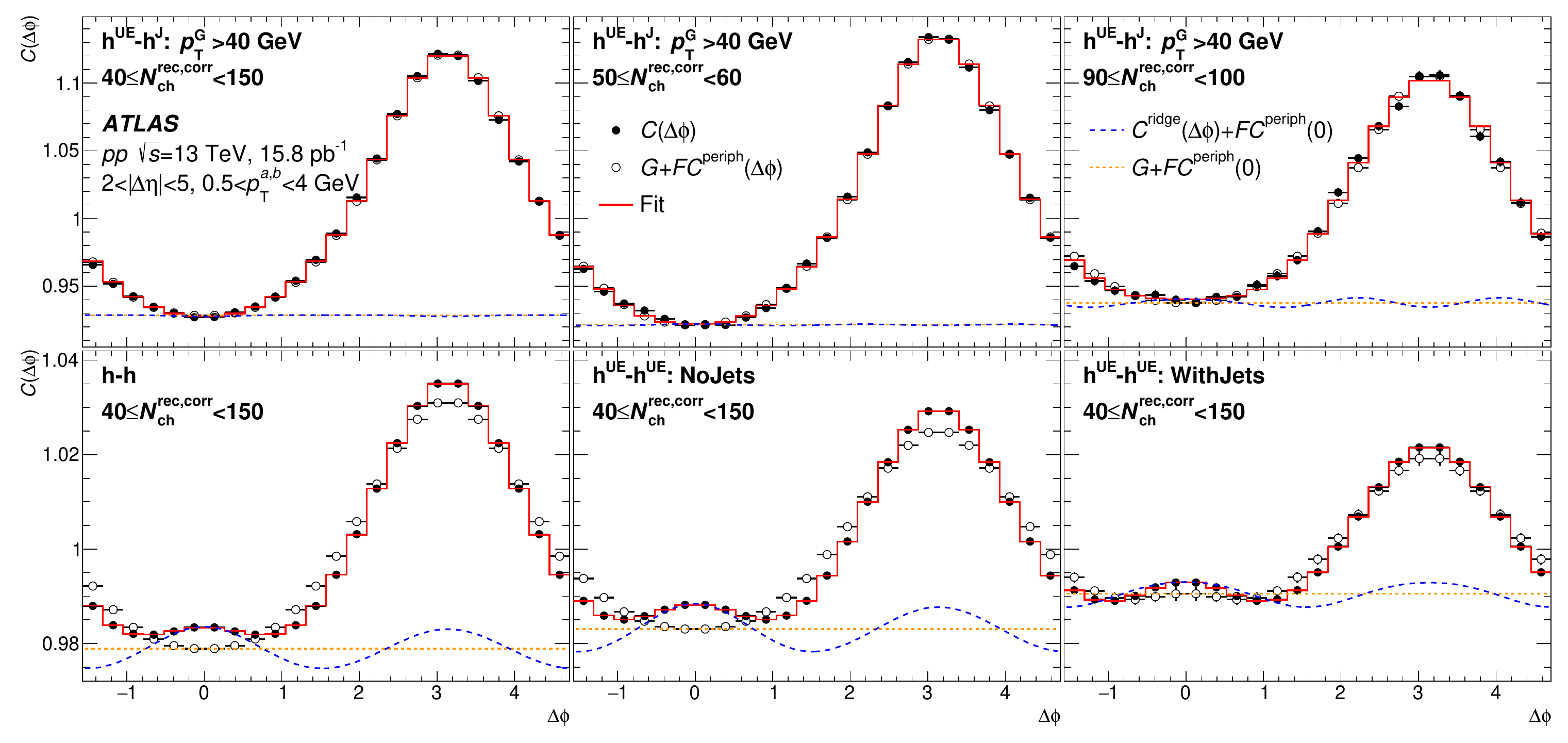}
\caption{
Template-fits to the two-particle correlations in $\dphi$.
Events with $10\le\ntrkreccorr<30$ are used as the peripheral reference.
The solid points indicate the measured 2PC, the open circles show the scaled
and shifted peripheral reference, and the continuous line shows the fit.
The dashed line shows the second-order harmonic component, and the dotted line
shows the pedestal of the fit shifted up by $FC^{\mathrm{periph}}(0)$.
The top row corresponds to different multiplicity intervals for the \hjet\ class.
The left, center and right panels in the bottom row correspond
to the \hh, \NoJet, and \WithJet\ classes, respectively,
for the 40--150 multiplicity interval.
}
\label{fig:templateFitResults}
\end{figure}

Figure~\ref{fig:templateFitResults} compares the \tpcs\ for all classes,
except the \All\ class.
The figure also shows the template fits including the components
of the fits.
In general, the template fits describe the \tpcs\ quite well.
A near-side ridge is visible for the \hh, \WithJet\ and \NoJet\ cases,
while the $\cperi(\dphi)$ appears to describe the full distribution in the \hjet\ case.
 
\begin{figure}[h]
\centering
\includegraphics[width=0.5\linewidth]{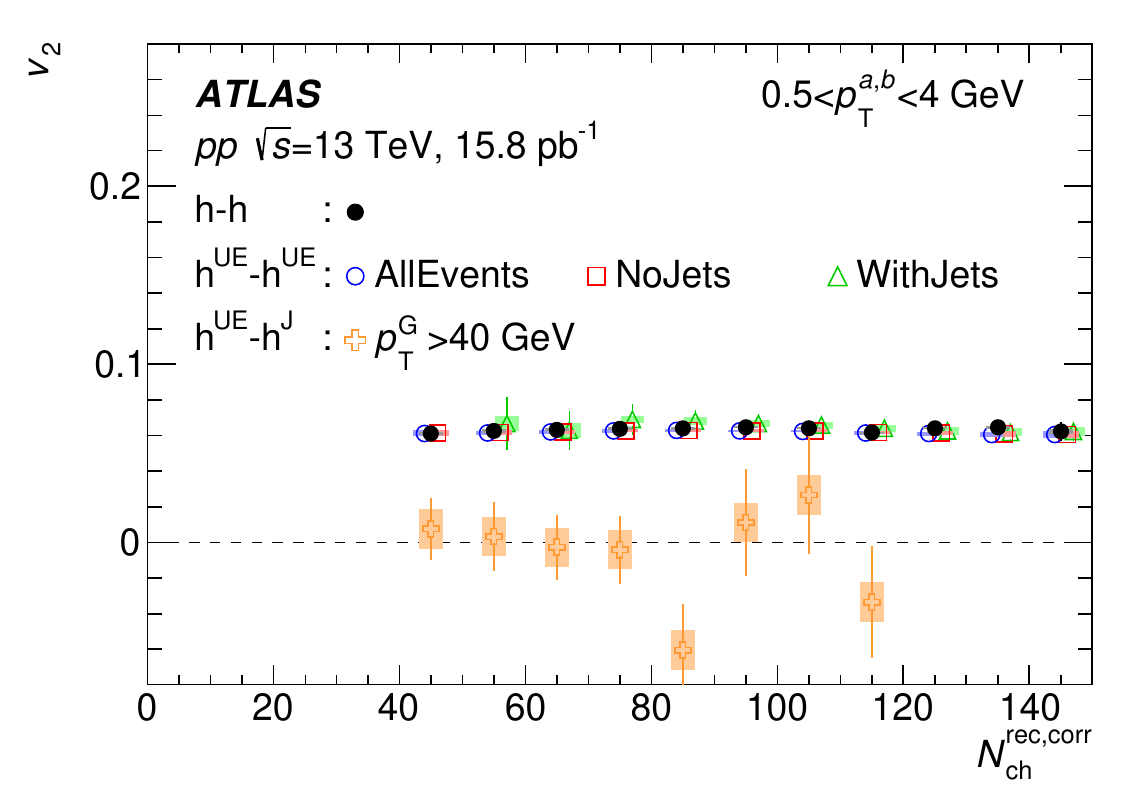}
\caption{
The multiplicity dependence of $v_2$ for $2<|\deta|<5$.
Events with $10\le\ntrkreccorr<30$ are used as the peripheral reference.
Jets with $\ptp>15~\GeV$ are used to classify the \NoJet\ and \WithJet\ samples.
The data point for the \WithJet\ case has a particularly large statistical uncertainty
in the 40--50 multiplicity interval and is not shown.
The data-points for the \All, \NoJet, and \WithJet\
samples are slightly shifted along the $x$-axis for clarity.
The error bars and bands correspond to statistical and systematic
uncertainties, respectively.
}
\label{fig:multDeptV2}
\end{figure}

\begin{figure}[h]
\centering
\includegraphics[width=0.5\linewidth]{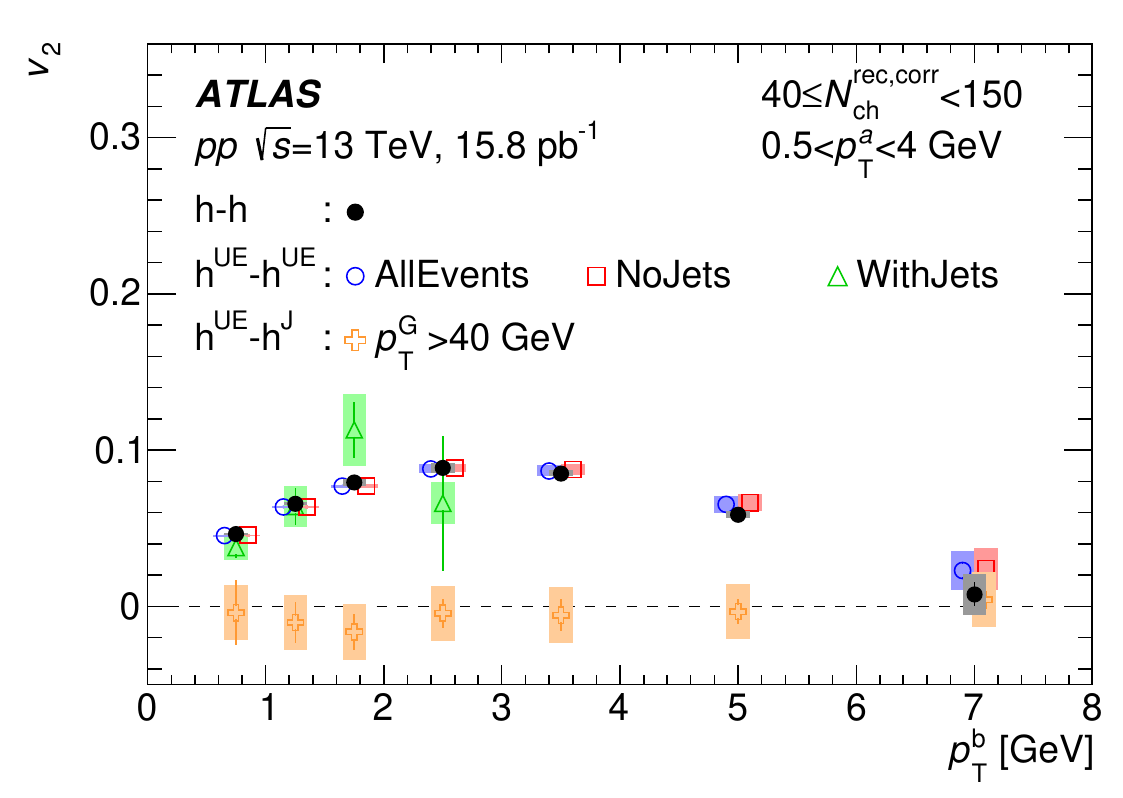}
\caption{
The \ptb\ dependence of the $v_2$ obtained for the 40--150 multiplicity interval for $2<|\deta|<5$.
Events with $10\le\ntrkreccorr<30$ are used as the peripheral reference.
Jets with $\ptp>15~\GeV$ are used to classify the \NoJet\ and \WithJet\ samples.
The data-points for the \hh\ sample are drawn at the nominal values
while the data-points for the \All, \NoJet, and the highest \ptb\
point of \hjet\ samples are shifted slightly for clarity.
The error bars and bands correspond to statistical and systematic
uncertainties, respectively.
}
\label{fig:ptDeptV2}
\end{figure}

Figure~\ref{fig:multDeptV2} shows the multiplicity dependence
of the $v_2$ for all five \tpc\ classes.
The $v_2$ values for the \hh\ case vary weakly with multiplicity,
as previously reported in Refs.~\cite{HION-2015-09,HION-2016-01}.
The $v_2$ values in the \All, \NoJet, and \WithJet\ cases,
are all consistent with the \hh\ result.
This demonstrates that removing tracks associated with jets does not
impact the long-range UE correlations, and nor does the presence (or absence) of
jets in an event.
Within uncertainties, the $v_2$ values in the \hjet\ case are consistent with zero.
The mean $v_2$ for the \hjet\ correlations over the 40--150 multiplicity range is
$-0.009 \pm 0.010(\mathrm{statistical}) \pm 0.014 (\mathrm{systematic})$.
This indicates that particles produced in hard scattering processes (with $\ptp>40$~\GeV) do not
contribute significantly to the long-range correlation observed in \pp\ collisions.
Figure~\ref{fig:ptDeptV2} shows the \pt-dependence of the $v_2$.
The differential $v_2(\pt)$ values in the \All, \NoJet\ and \WithJet\
cases are found to be consistent with the \hh\ case.
Again, within uncertainties, the \hjet\ $v_2$ values are consistent with zero,
across the entire measured \pt\ range.
The findings drawn from the \pt\ dependence are consistent with those from the
multiplicity dependence, and similarly demonstrate that the presence or absence
of jets has no influence on the flow of the UE and that there
are no correlations between jet-fragments and the UE\@.
The features of the $v_2$ values discussed above do not show
any systematic variation with the jet selections,
for example the \ptp\ thresholds used in the analysis,
as discussed in Ref.~\cite{SupplementaryMaterial}.

\FloatBarrier

In conclusion, this \papertype\ studies long-range \tpcs\ in \pp\ collisions
when rejecting tracks in the vicinity of jets, and the correlations
between jet constituent tracks and tracks from the UE\@.
The \tpcs\ are analyzed using a template-fit procedure, previously developed
by ATLAS~\cite{HION-2016-01}, which extracts second-order Fourier coefficients
($v_2$) of the anisotropy.
These results demonstrate that the magnitude of the $v_2$ is not affected
when removing tracks associated with jets,
or by the presence or absence of jets in the event.
The $v_2$ measured with correlations between jet constituents with
$\pt<8~\GeV$ and UE tracks are consistent with zero within uncertainties.
These features are observed both in the $v_2$ multiplicity and \pt\ dependence.
 
The observation that fragments of high-\pT jets in \pp\ collisions
do not have measurable long-range azimuthal correlations with the UE\@,
and that the production of $Z$ bosons~\cite{HION-2017-11} or jets does not
significantly influence the long-range correlations between UE particles,
suggest a complete ``factorization'' between hard-scattering
processes and the physics responsible for the ridge.
Further studies are needed to extend this measurement to higher \pt\
to compare with previous measurements in \pPb\
collisions~\cite{HION-2018-20} where such factorization is broken.
This \papertype\ provides important insights into the origin of the
long-range correlations observed in \pp\ collisions
and offers new fundamental input to theoretical models.

\section*{Acknowledgements}


We thank CERN for the very successful operation of the LHC, as well as the
support staff from our institutions without whom ATLAS could not be
operated efficiently.
 
We acknowledge the support of
ANPCyT, Argentina;
YerPhI, Armenia;
ARC, Australia;
BMWFW and FWF, Austria;
ANAS, Azerbaijan;
CNPq and FAPESP, Brazil;
NSERC, NRC and CFI, Canada;
CERN;
ANID, Chile;
CAS, MOST and NSFC, China;
Minciencias, Colombia;
MEYS CR, Czech Republic;
DNRF and DNSRC, Denmark;
IN2P3-CNRS and CEA-DRF/IRFU, France;
SRNSFG, Georgia;
BMBF, HGF and MPG, Germany;
GSRI, Greece;
RGC and Hong Kong SAR, China;
ISF and Benoziyo Center, Israel;
INFN, Italy;
MEXT and JSPS, Japan;
CNRST, Morocco;
NWO, Netherlands;
RCN, Norway;
MEiN, Poland;
FCT, Portugal;
MNE/IFA, Romania;
MESTD, Serbia;
MSSR, Slovakia;
ARRS and MIZ\v{S}, Slovenia;
DSI/NRF, South Africa;
MICINN, Spain;
SRC and Wallenberg Foundation, Sweden;
SERI, SNSF and Cantons of Bern and Geneva, Switzerland;
MOST, Taiwan;
TENMAK, T\"urkiye;
STFC, United Kingdom;
DOE and NSF, United States of America.
In addition, individual groups and members have received support from
BCKDF, CANARIE, Compute Canada and CRC, Canada;
PRIMUS 21/SCI/017 and UNCE SCI/013, Czech Republic;
COST, ERC, ERDF, Horizon 2020 and Marie Sk{\l}odowska-Curie Actions, European Union;
Investissements d'Avenir Labex, Investissements d'Avenir Idex and ANR, France;
DFG and AvH Foundation, Germany;
Herakleitos, Thales and Aristeia programmes co-financed by EU-ESF and the Greek NSRF, Greece;
BSF-NSF and MINERVA, Israel;
Norwegian Financial Mechanism 2014-2021, Norway;
NCN and NAWA, Poland;
La Caixa Banking Foundation, CERCA Programme Generalitat de Catalunya and PROMETEO and GenT Programmes Generalitat Valenciana, Spain;
G\"{o}ran Gustafssons Stiftelse, Sweden;
The Royal Society and Leverhulme Trust, United Kingdom.
 
The crucial computing support from all WLCG partners is acknowledged gratefully, in particular from CERN, the ATLAS Tier-1 facilities at TRIUMF (Canada), NDGF (Denmark, Norway, Sweden), CC-IN2P3 (France), KIT/GridKA (Germany), INFN-CNAF (Italy), NL-T1 (Netherlands), PIC (Spain), ASGC (Taiwan), RAL (UK) and BNL (USA), the Tier-2 facilities worldwide and large non-WLCG resource providers. Major contributors of computing resources are listed in Ref.~\cite{ATL-SOFT-PUB-2021-003}.


\printbibliography

\clearpage
\appendix
\part*{Appendix}
\addcontentsline{toc}{part}{Appendix}

\section{Supplemental Material Allowed by Journal to Accompany Online Version of Paper}

\subsection{Detailed trigger description}
Multiple triggers were used to record the data
and can broadly be classified into three categories.
The first category of triggers selects a set of minimum-bias events through
a L1 trigger that requires a signal in at least one MBTS counter,
and a second trigger that requires at least one reconstructed track
with $\pT>0.2~\GeV$ at the HLT\@.
The second category is a set of high-multiplicity triggers that apply a L1 requirement on
either the transverse energy (\et) in the calorimeters or a hit in at least
one MBTS counter on each side,
and an HLT requirement on the multiplicity of HLT-reconstructed tracks
with $\pT>0.4~\GeV$ that is associated with the reconstructed vertex with
the highest multiplicity in the event.
The third is a set of triggers that enhance the rate of jets.
These include a set of triggers that require a jet at L1 and at the HLT,
and another set of triggers that require a minimum threshold on
the total \et\ in the calorimeter at L1.

\subsection{Detailed track selection criteria}
The criteria used to select tracks include the requirements of $\pT > 0.4$~\GeV\ and $|\eta|<2.5$,
a hit in the IBL or a hit in the pixel layer next to the IBL,
and a minimum of six hits in the SCT\@.
Additionally, the transverse impact parameter of the track relative to the
average beam position,
and the longitudinal impact parameter of the track relative to the vertex
are both required to be less than 1.5~mm.
To remove tracks with mis-measured \pt\ due to interactions with the
material or other effects, the track-fit $\chi^2$ probability is required to be
larger than 0.01 for tracks having $\pT > 10$~\GeV.

\subsection{Multiplicity distributions}
\begin{figure}[h]
\centering
\includegraphics[width=0.6\linewidth]{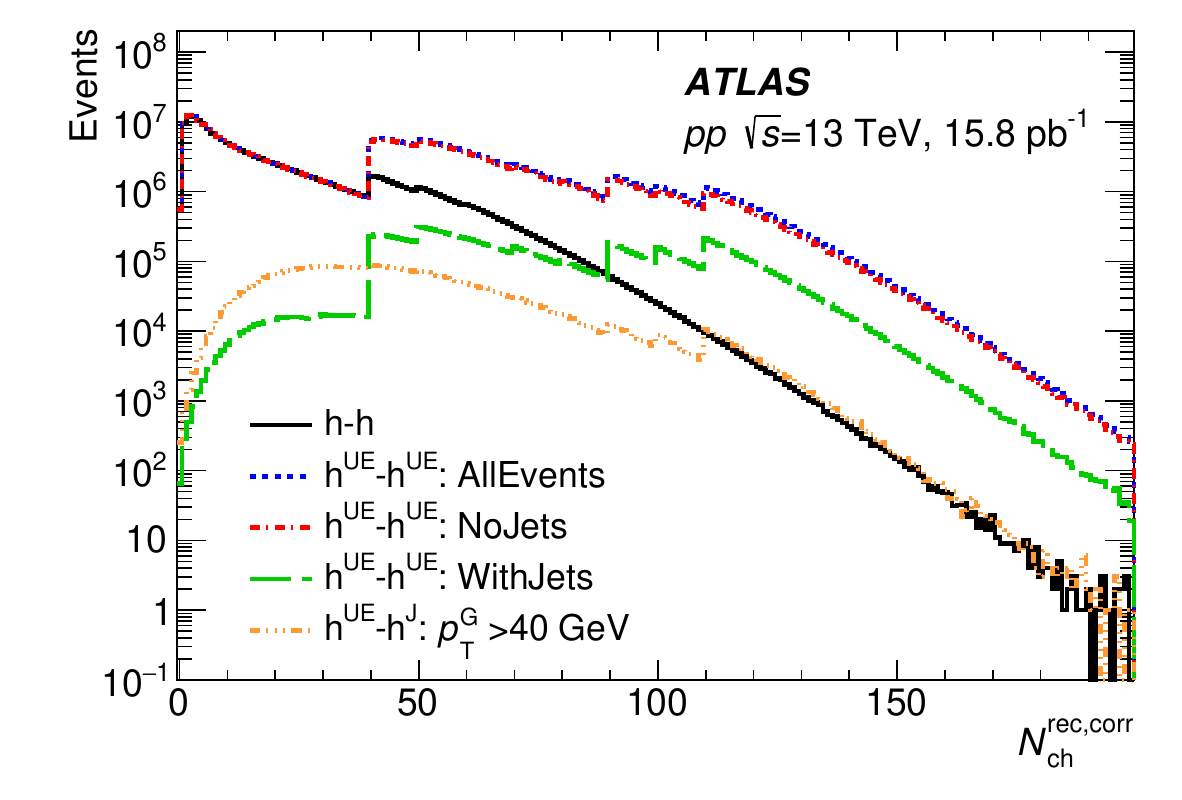}
\caption{
The distribution of \ntrkreccorr\ for the $\sqs=13$~\TeV\ \pp\ data used in this analysis.
The different lines denote the \hh, \All, \NoJet, \WithJet, and \hjet\ cases.
The discontinuities in the distributions correspond to different high-multiplicity trigger thresholds.
}
\label{fig:ntrk_5case}
\end{figure}

Figure~\ref{fig:ntrk_5case} shows the distribution of \ntrkreccorr\ for
the events in the \hh, \All, \NoJet, \WithJet, and \hjet\ cases.
The default peripheral reference in the template-fits is constructed using
the events in the 10--30 multiplicity interval.
The average minimum-bias multiplicity of reconstructed charged-particle tracks
in 13~\TeV\ \pp\ collisions in ATLAS is $\sim20$,
and the default range for the peripheral reference
is taken to be a $\pm10$ window around this mean value.
Alternative peripheral references built from the 10--40 and 20--40
multiplicity intervals are used to evaluate systematic uncertainties,
where the lower and upper multiplicity ranges are increased by 10
from the default interval.
The 0--10 multiplicity interval, is excluded from the peripheral references as
1) the \WithJet\ and \hjet\ cases have very few events at these low multiplicities,
and 2) such low multiplicity are likely to contain a significant
contribution from diffractive events.
 
\subsection{\PYTHIA~8 Embedding}
This section motivates the choice of the 4~\GeV\ requirement used
in Eq.~\eqref{eq:ptp} to remove the bias from the UE modulation on
the jet selection.
The effect of UE modulation on the jet selection is estimated by a
toy \PYTHIA~8 embedding study.
The \PYTHIA~8 events are simulated using the Monash 2013 tune with
multi-parton interaction off and initial-state radiation on.
Jet reconstruction is performed on these generated events,
using the anti-$k_{\mathrm{T}}$ clustering algorithm with a radius parameter of 0.4.
The jets thus produced are called \textit{generated-jets} hereafter.
The generated events are filtered by requiring the events to have a generated-jet
with \pt\ greater than 15~\GeV and a balanced generated-jet with
$\pt>10~\GeV$ and $|\dphi| > 5\pi/6$ relative to the first jet.
The generated (stable\footnote{The generated particles that are not decayed further by \PYTHIA~8 are called stable here.})
particles are embedded onto minimum bias data events.
After the embedding, jets are reconstructed using the embedded particles and
the original particle flow objects~\cite{PERF-2015-09} present in the data-event, using the anti-$k_{\mathrm{T}}$
clustering algorithm with a radius parameter of 0.4.
These jets are called \textit{embedded-jets} hereafter.
Similar jet clustering is also done using only the particle flow objects in the
minimum-bias data used in the embedding study.
These jets are called \textit{data-jets} hereafter.

\begin{figure}[h]
\centering
\includegraphics[width=0.95\linewidth]{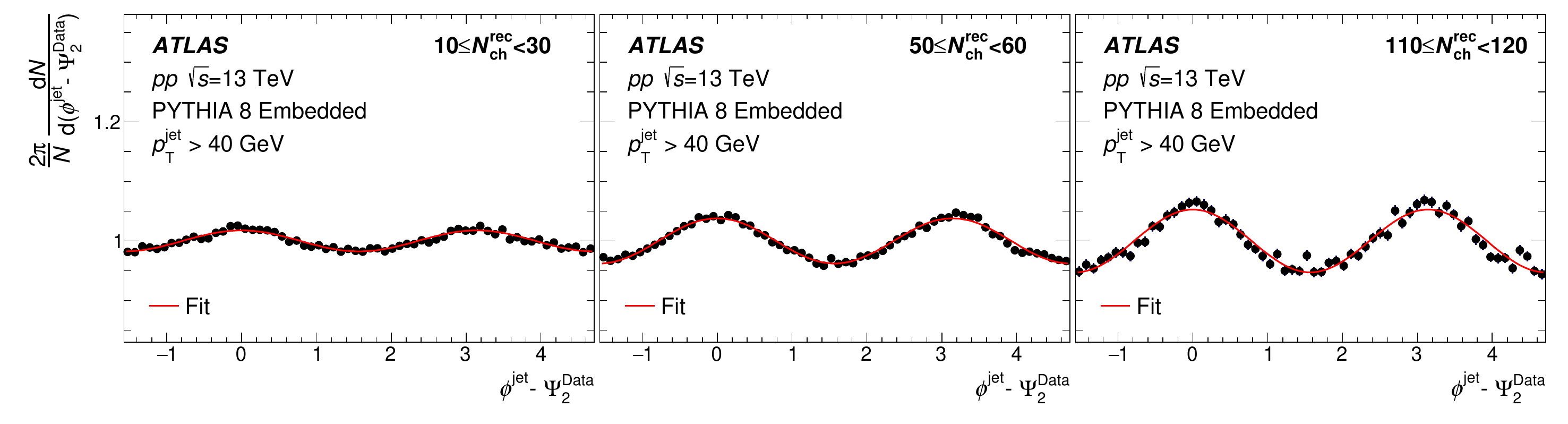}
\caption{
The distribution of $\phi^{\mathrm{jet}}-\Psi_2^{\mathrm{Data}}$
for the embedded jets.
The continuous line indicates a Fourier fit to the distribution that
includes a $2^{\mathrm{nd}}$ order modulation.
The different panels correspond to different multiplicity intervals.
}
\label{fig:jetv2_oldPt}
\end{figure}

The embedded-jets are required to match a generated-jet with
$\Delta R=\sqrt{\deta^2+\dphi^2}<0.1$.
The embedded-jets are also required to be separated from
data-jets by $\Delta R>0.8$.
The \pt\ of the embedded-jets is always larger than the \pt\ of the
matched generated-jet as the particle flow objects in the data always push up the \pt.
The issue of concern for this analysis is that this increase
in the \pt\ is dependent on the azimuthal angle that the jet makes with the
second-order event-plane, due to the modulation in the UE\@.
To visualize the bias from the UE modulation, the difference
between the azimuthal angle of the embedded-jet, $\phi^{\mathrm{jet}}$,
and the second-order event plane angle, $\Psi_2^{\mathrm{Data}}$, is plotted
in Figure~\ref{fig:jetv2_oldPt}.
The $\Psi_2$ angle is calculated using particles from the data events only,
excluding particles within one unit in $\eta$ around any data-jets with
\pt\ greater than 15~\GeV.
Since \PYTHIA~8 events are uncorrelated with the data event onto which they are embedded,
the distribution of $\phi^{\mathrm{jet}} - \Psi_2^{\mathrm{Data}}$ is,
by construction, constant for the generated-jets.
However, for the embedded-jets the corresponding distribution is modulated, as shown in
Figure~\ref{fig:jetv2_oldPt}.
 
\begin{figure}[h]
\centering
\includegraphics[width=0.95\linewidth]{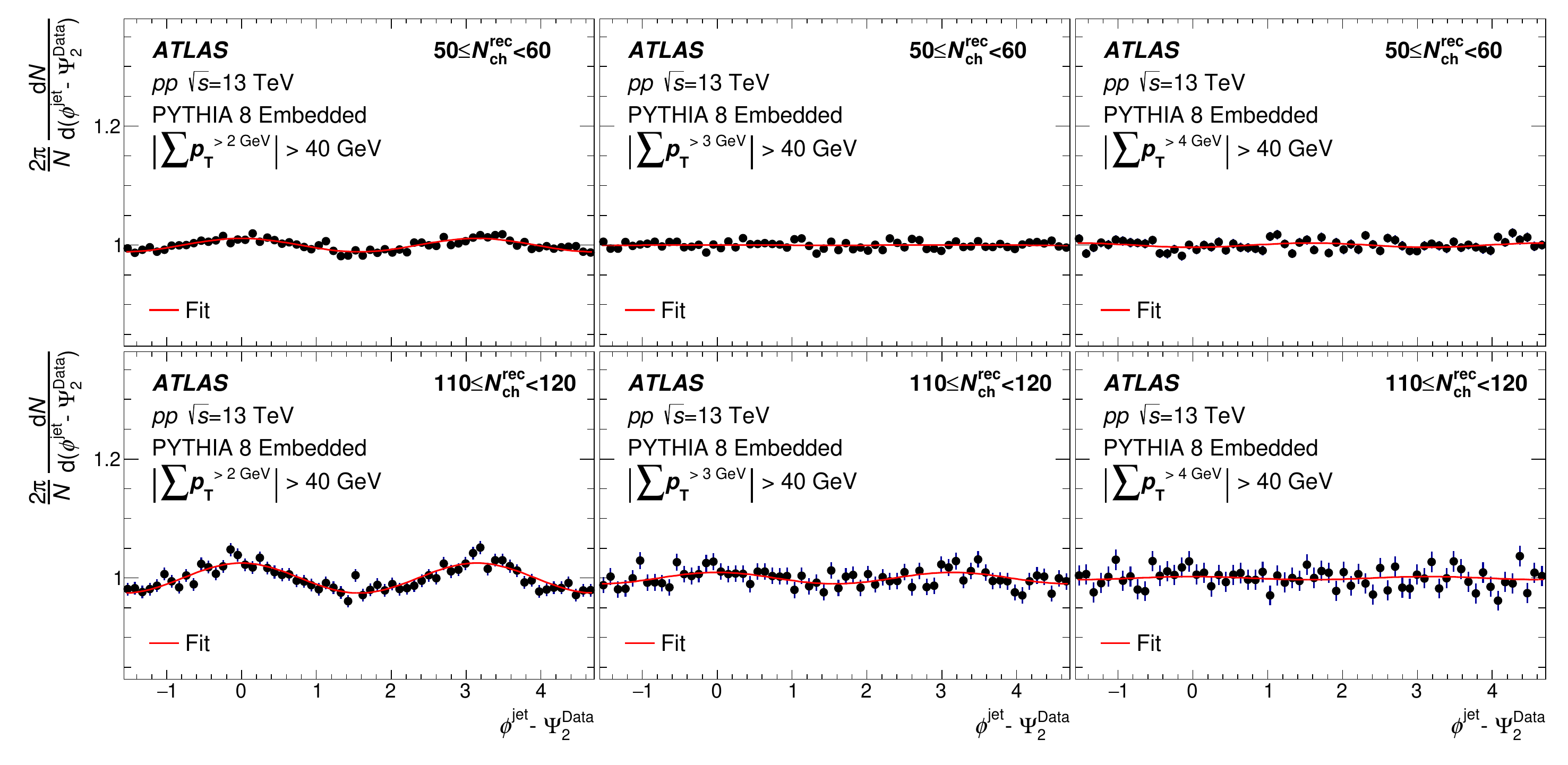}
\caption{
The distribution of $\phi^{\mathrm{jet}}-\Psi_2^{\mathrm{Data}}$
for the embedded jets.
The continuous line indicates a Fourier fit to the distribution that
includes a $2^{\mathrm{nd}}$ order modulation.
From left to right, the three panels in each row correspond to increasing
thresholds $X$ (in Eq.~\eqref{eq:ptp_aux}) of 2~\GeV, 3~\GeV, and 4~\GeV.
The top (bottom) row corresponds to the 50--60 (110--120) multiplicity interval.
}
\label{fig:jetv2_newPt}
\end{figure}

Since the modulation effects are dominated by low \pt\ particles,
grooming the jets to remove soft particles can remove the UE bias.
A groomed jet \pt\ definition is proposed here, by summing the \pt\ of jet constituents
with \pt\ greater than a particular threshold $X$:
\begin{eqnarray}
p_{\mathrm{T}}^{\mathrm{G}}=\left|\sum{\boldsymbol{p}_{\mathrm{T}}^{\ >\ \mathrm{X}\ \GeV}}\right|,
\label{eq:ptp_aux}
\end{eqnarray}
Figure~\ref{fig:jetv2_newPt} shows the $\phi^{\mathrm{jet}} - \Psi_2^{\mathrm{Data}}$
using different values of the threshold $X$ in Eq.~\eqref{eq:ptp_aux}.
The modulation in the distribution systematically decreases with increasing $X$,
and is nearly gone for $X=4$~\GeV.
 
\subsection{Illustration of the procedure for selecting $\tue-\tue$ and \hjet\ pairs}
 
\begin{figure}[h]
\centering
\includegraphics[width=0.5\linewidth]{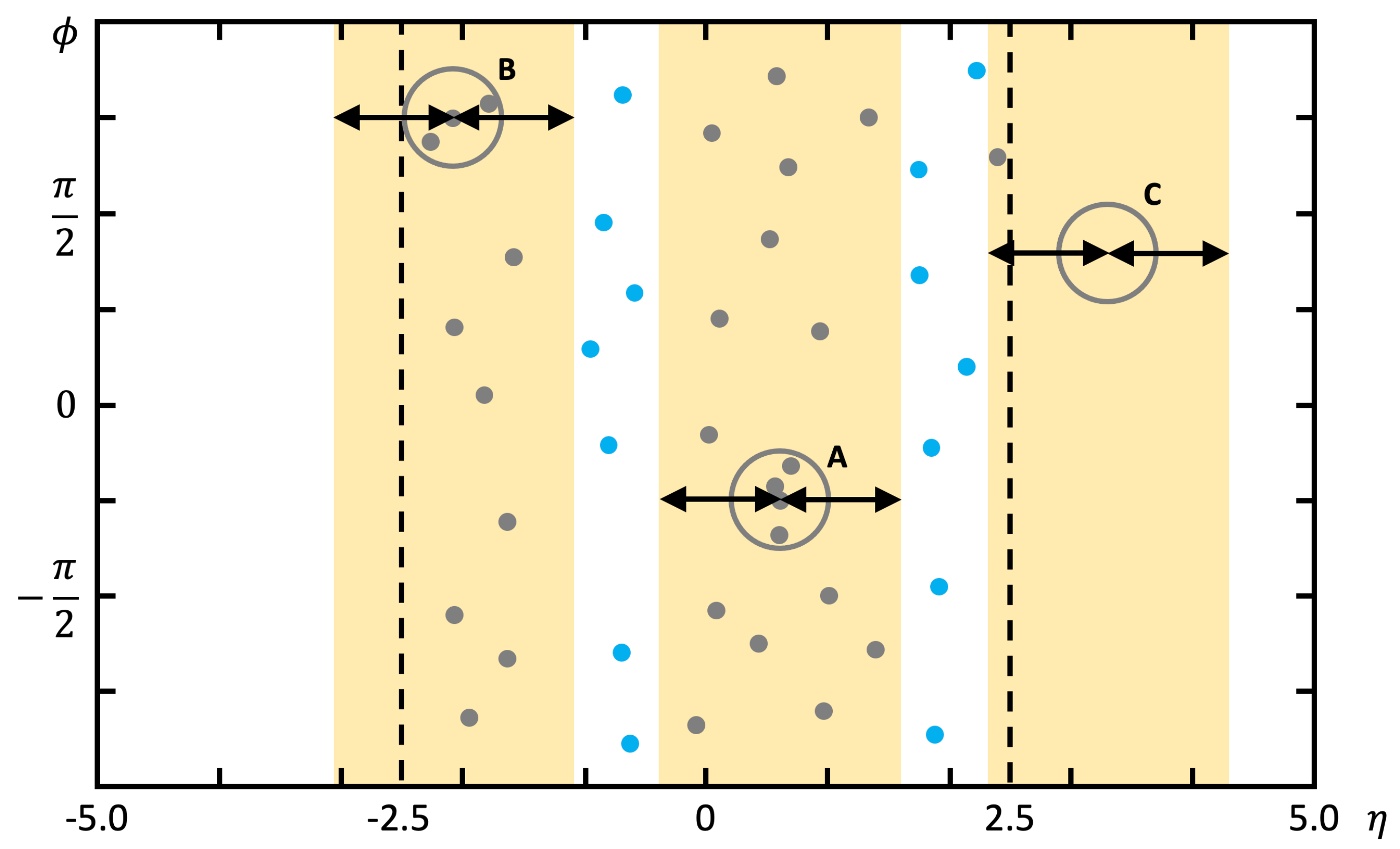}
\caption{
Conceptual depiction of the procedure to measure the $\tue-\tue$ correlations.
The $x$ and $y$-axes depict $\eta$ and $\phi$, respectively.
The ATLAS calorimeter system extends over $\pm$4.9 in $\eta$,
over which range the jets are reconstructed.
The vertical dashed lines at $\eta=\pm2.5$ indicate the acceptance
of the ATLAS ID, over which charged-particle tracks are reconstructed.
The figure shows an event with three $\ptp > 15~\GeV$ jets (grey circles).
The vertical yellow bands indicate a $\pm1$ $\eta$ window around the jets.
The reconstructed tracks within the yellow bands (grey dots) are excluded from
the $\tue-\tue$ correlation study.
The blue dots represent the tracks that are used in the
$\tue-\tue$ correlation.
}
\label{fig:cartoon_ueue}
\end{figure}
 
\begin{figure}[h]
\centering
\includegraphics[width=0.5\linewidth]{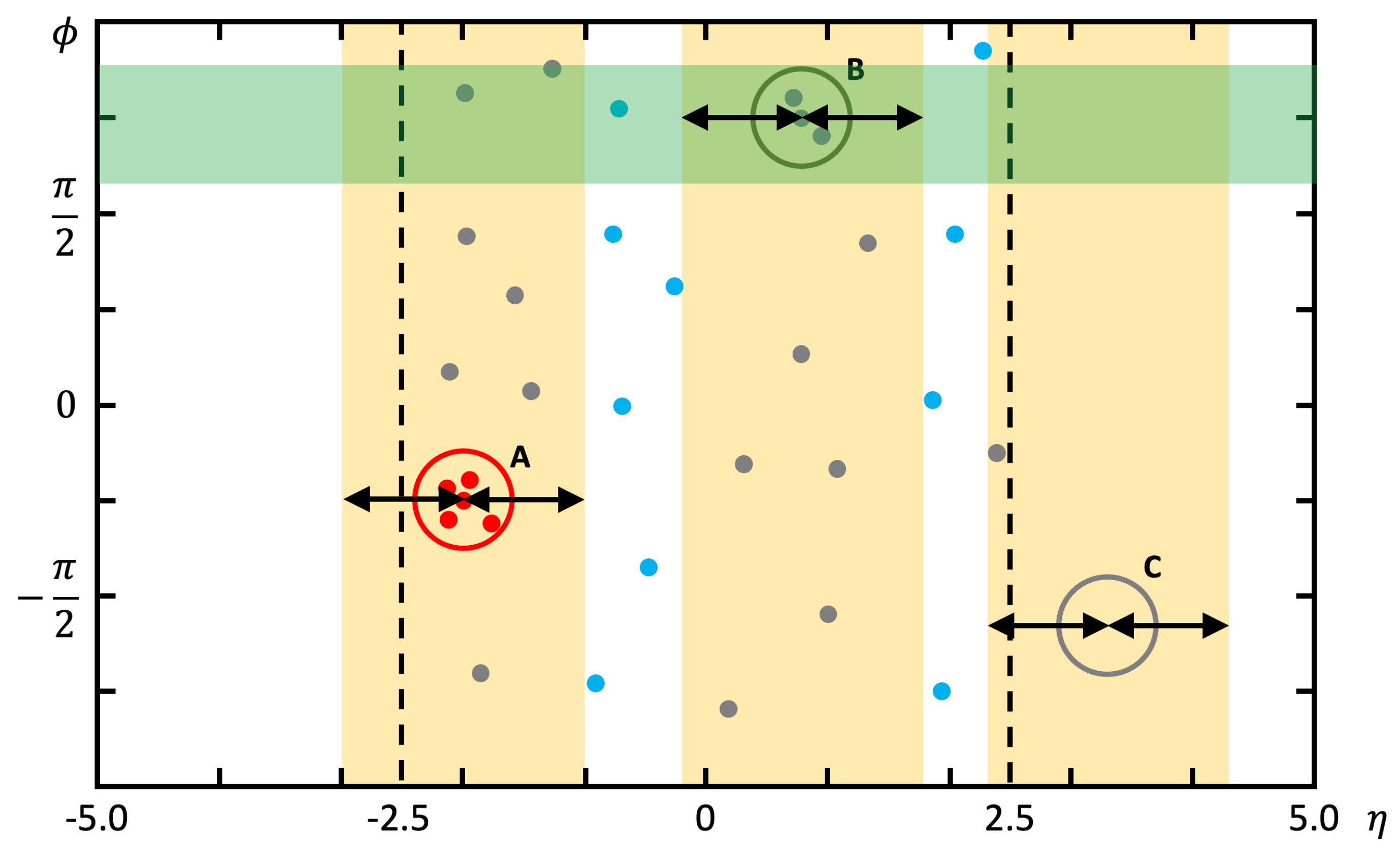}
\caption{
Conceptual depiction of the procedure to measure the \hjet\ correlations.
The $x$ and $y$-axes depict $\eta$ and $\phi$, respectively.
The ATLAS calorimeter system extends over $\pm$4.9 in $\eta$,
over which range the jets are reconstructed.
The vertical dashed lines at $\eta=\pm2.5$ indicate the acceptance
of the ATLAS ID, over which charged-particle tracks are reconstructed.
The figure shows an event with two $\ptp > 15~\GeV$ jets (grey circles,
labelled as $B$ and $C$),
and a $\ptp>40~\GeV$ jet (red circle, labelled as $A$).
The red dots represent the tracks that are considered as jet constituents (\tjet)
of the $\ptp>40~\GeV$ jet.
The vertical yellow bands indicate a $\pm1$ $\eta$ window around the jets.
The blue dots represent the tracks that are treated as
\UE\ tracks (see Figure~\ref{fig:cartoon_ueue}).
The green band indicates the region scanned to find the
balancing jet for jet-$A$.
In this case the balancing jet is jet-$B$.
In the absence of a balancing jet, the event is not used for the \hjet\ correlation.
}
\label{fig:cartoon_jetue}
\end{figure}

Figures~\ref{fig:cartoon_ueue} and~\ref{fig:cartoon_jetue} illustrate
the procedure for selecting the $\tue-\tue$ and \hjet\ pairs, respectively.

\subsection{Cross checks on jet selections used in the analysis}
This section discusses several checks that are made to test the sensitivity of the results
on the jet selections used in the analysis.
\begin{itemize}
\item
The sensitivity of the \hjet\ $v_2$ measurements to the \ptp\ selection
is evaluated by varying the \ptp\ selection threshold from the nominal value
of 40~\GeV\ to 35~\GeV\ and 50~\GeV.
\item
The sensitivity of the $v_2$ to the \pt\ threshold applied on the
jet constituents in Eq.~\eqref{eq:ptp} is checked by raising it from
its default value of 4~\GeV\ to 4.5~\GeV.
\item
A threshold of \ptp\ = 15 \GeV\ is applied to separate the hard and soft
processes in the event.
This threshold is an analysis choice as there is no \pt\ value that fully
separates hard and soft processes.
The sensitivity of the results to the choice of the jet \ptp\ is investigated
by increasing the threshold from its nominal value of 15 GeV to 20 GeV.
\item
As stated in the \papertype\,
as isolation requirement is imposed on the $\ptp>40$~\GeV\ jets
in the \hjet\ correlations.
This isolation requirement requires that there are no $\ptp>15$~\GeV\
jets within $\Delta R=1$ of the $\ptp>40$~\GeV\
jets used in the \hjet\ correlation analysis.
As a cross-check the measurements are repeated with this isolation
requirement removed.
\end{itemize}
For all these cases no significant variation is observed
in the measurements, and the results with these variations
are consistent with the nominal ones within the quoted
systematic and statistical uncertainties.

\clearpage
 
\begin{flushleft}
\hypersetup{urlcolor=black}
{\Large The ATLAS Collaboration}

\bigskip

\AtlasOrcid[0000-0002-6665-4934]{G.~Aad}$^\textrm{\scriptsize 102}$,
\AtlasOrcid[0000-0002-5888-2734]{B.~Abbott}$^\textrm{\scriptsize 120}$,
\AtlasOrcid[0000-0002-1002-1652]{K.~Abeling}$^\textrm{\scriptsize 55}$,
\AtlasOrcid[0000-0001-5763-2760]{N.J.~Abicht}$^\textrm{\scriptsize 49}$,
\AtlasOrcid[0000-0002-8496-9294]{S.H.~Abidi}$^\textrm{\scriptsize 29}$,
\AtlasOrcid[0000-0002-9987-2292]{A.~Aboulhorma}$^\textrm{\scriptsize 35e}$,
\AtlasOrcid[0000-0001-5329-6640]{H.~Abramowicz}$^\textrm{\scriptsize 151}$,
\AtlasOrcid[0000-0002-1599-2896]{H.~Abreu}$^\textrm{\scriptsize 150}$,
\AtlasOrcid[0000-0003-0403-3697]{Y.~Abulaiti}$^\textrm{\scriptsize 117}$,
\AtlasOrcid[0000-0003-0762-7204]{A.C.~Abusleme~Hoffman}$^\textrm{\scriptsize 137a}$
% ...
\bigskip
\\

% ...
$^{29}$Physics Department, Brookhaven National Laboratory, Upton NY; United States of America.\\
% ...
$^{35}$$^{(a)}$Facult\'e des Sciences Ain Chock, R\'eseau Universitaire de Physique des Hautes Energies - Universit\'e Hassan II, Casablanca;$^{(b)}$Facult\'{e} des Sciences, Universit\'{e} Ibn-Tofail, K\'{e}nitra;$^{(c)}$Facult\'e des Sciences Semlalia, Universit\'e Cadi Ayyad, LPHEA-Marrakech;$^{(d)}$LPMR, Facult\'e des Sciences, Universit\'e Mohamed Premier, Oujda;$^{(e)}$Facult\'e des sciences, Universit\'e Mohammed V, Rabat;$^{(f)}$Institute of Applied Physics, Mohammed VI Polytechnic University, Ben Guerir; Morocco.\\
% ...
$^{49}$Fakult\"{a}t Physik , Technische Universit{\"a}t Dortmund, Dortmund; Germany.\\
% ...
$^{55}$II. Physikalisches Institut, Georg-August-Universit\"{a}t G\"ottingen, G\"ottingen; Germany.\\
% ...
$^{102}$CPPM, Aix-Marseille Universit\'e, CNRS/IN2P3, Marseille; France.\\
% ...
$^{117}$Department of Physics, New York University, New York NY; United States of America.\\
% ...
$^{120}$Homer L. Dodge Department of Physics and Astronomy, University of Oklahoma, Norman OK; United States of America.\\
% ...
$^{137}$$^{(a)}$Departamento de F\'isica, Pontificia Universidad Cat\'olica de Chile, Santiago;$^{(b)}$Millennium Institute for Subatomic physics at high energy frontier (SAPHIR), Santiago;$^{(c)}$Instituto de Investigaci\'on Multidisciplinario en Ciencia y Tecnolog\'ia, y Departamento de F\'isica, Universidad de La Serena;$^{(d)}$Universidad Andres Bello, Department of Physics, Santiago;$^{(e)}$Instituto de Alta Investigaci\'on, Universidad de Tarapac\'a, Arica;$^{(f)}$Departamento de F\'isica, Universidad T\'ecnica Federico Santa Mar\'ia, Valpara\'iso; Chile.\\
% ...
$^{150}$Department of Physics, Technion, Israel Institute of Technology, Haifa; Israel.\\
$^{151}$Raymond and Beverly Sackler School of Physics and Astronomy, Tel Aviv University, Tel Aviv; Israel.\\
% ...
$^{a}$ Also Affiliated with an institute covered by a cooperation agreement with CERN.\\
$^{b}$ Also at An-Najah National University, Nablus; Palestine.\\
$^{c}$ Also at APC, Universit\'e Paris Cit\'e, CNRS/IN2P3, Paris; France.\\
$^{d}$ Also at Borough of Manhattan Community College, City University of New York, New York NY; United States of America.\\
$^{e}$ Also at Center for High Energy Physics, Peking University; China.\\
$^{f}$ Also at Center for Interdisciplinary Research and Innovation (CIRI-AUTH), Thessaloniki ; Greece.\\
% ...

\end{flushleft}


\end{document}